\documentclass[12pt,a4paper,final]{iopart}

\usepackage{iopams}  
\usepackage{graphicx}

\expandafter\let\csname equation*\endcsname\relax
\expandafter\let\csname endequation*\endcsname\relax

\usepackage{mathtools}
\usepackage{amsmath}
\usepackage{amssymb}
\usepackage[breaklinks=true,colorlinks=true,linkcolor=blue,urlcolor=blue,citecolor=blue]{hyperref}

\begin{document}

\title{Scaling Properties of Superoscillations and the Extension to Periodic Signals}

\author{Eugene Tang}
\address{Division of Physics,
Mathematics and Astronomy, California Institute of Technology, Pasadena, CA 91125, USA}

\author{Lovneesh Garg}
\address{Department of Applied Mathematics, University of Waterloo, Waterloo, Canada}

\author{Achim Kempf}
\address{Departments of Applied Mathematics and Physics, University of Waterloo, Waterloo, Canada}

\begin{abstract}
Superoscillatory wave forms, i.e., waves that locally oscillate faster than their highest Fourier component, possess unusual properties that make them of great interest from quantum mechanics to signal processing. However, the more pronounced the desired superoscillatory behavior is to be, the more difficult it becomes to produce, or even only calculate, such highly fine-tuned wave forms in practice. Here, we investigate how this sensitivity to preparation errors scales for a method for constructing superoscillatory functions which is optimal in the sense that it minimizes the energetic expense. We thereby also arrive at very accurate approximations of functions which are so highly superoscillatory that they cannot be calculated numerically. 
We then investigate to what extent the scaling and sensitivity results for superoscillatory functions on the real line extend to the experimentally important case of superoscillatory functions that are periodic. 
\end{abstract}


\section{Introduction}
It used to be thought that functions $f(t)$ which are bandlimited to a frequency $\Omega$ cannot exhibit local oscillations with frequencies larger than $\Omega$. Indeed, according to the Shannon-Nyquist sampling theorem, knowledge of the amplitudes $\{f(t_n)\}_{n=-\infty}^\infty$ of an $\Omega$-bandlimited function $f$ at a set of points with spacing $t_{n+1}-t_n = (2\Omega)^{-1}$ suffices to reconstruct the function everywhere. Intuitively, this suggests that local oscillations that are faster than this sample spacing would be missed and therefore should be absent from such functions. 

It was eventually found however, that this is not the case. In the early 1990s, Aharonov and Berry gave examples of bandlimited functions which exhibited arbitrarily rapid oscillations on a local stretch \cite{Aharonov1990, Berry1994a}. Such functions exhibit quite counter-intuitive behaviors and were termed \it superoscillations. \rm In hindsight, examples of superoscillatory behavior can be seen already in the works of Slepian et al. in the 1960s on the prolate spheroidal wave functions, a sequence of bandlimited functions which become superoscillatory \cite{Slepian,SlepianPollak}.

Over the past years, superoscillations have become of interest in several regards. Superoscillations have been shown to have unusual consequences in quantum physics, where, for example, a particle whose wave function is superoscillatory behaves as if ``spring loaded" when passing through a slit: if it is arranged that only the short-wavelength superoscillatory part of wave function passes through the slit, then the particle acquires a large predetermined increase in the expectation value of its transverse momentum \cite{Ferreira2004,calder} merely by passing through the slit. In addition, superoscillations arise in the context of quantum billiards \cite{Berry1994b} and weak measurements \cite{Berry2012}. Superoscillations have also been proposed to arise with the trans-Planckian problem of Hawking radiation \cite{Reznik1997, Beethoven}, and as a tool in the remote preparation of quantum states \cite{Reznik2015}. In the field of signal processing, superoscillations have been proposed as tools to achieve super-resolution \cite{Huang, Rogers, Lindberg, prain, BerryPopescu}. Also,  their unique properties push our understanding of information compression, with applications to temporal pulse compression beyond the Fourier limit \cite{Wong}.

While superoscillations possess a number of intriguing features, these do come with a cost. The more superoscillatory a function is, the larger the dynamic range of the function has to be \cite{Ferreira2004,ferreira2002,Ferreira2006}. Namely, if a function possesses superoscillations then the function will also possess stretches of oscillations whose amplitudes are much larger than the amplitudes of the superoscillations. Even moderately superoscillatory functions experience a dynamic range on the order of about $10^5$. This means that superoscillatory functions tend to be difficult to create or measure in practice, as the amplitudes of the superoscillating stretch will be very small compared to other amplitudes in the function.

For this reason, it is of paramount interest to construct superoscillatory functions so as to minimize their dynamic range. Equivalently, the task is to construct superoscillatory functions so that, after prescribing amplitudes of the superoscillations, the resulting superoscillatory function comes out as small as possible, say in the sense of possessing the minimum possible $L^2$ norm. The square of the $L^2$-norm is also known in the engineering literature as the energy of a signal. For simplicity, we will adopt this terminology here for all signals or wave functions. In this terminology, our aim here is to study the scaling behavior of those superoscillatory functions that require the least energy to pass through a finite number of points which are chosen to be oscillating faster than the highest frequency in the bandwidth. 

Since the initial discovery of superoscillations, several methods for constructing superoscillations have been proposed, for example, by shifting the zeros of bandlimited functions \cite{Qiao}, by uniformly approximating polynomials \cite{Chremmos}, or as the uniformly convergent limit of a sequence of functions \cite{Aharonov2011}. One of the very first methods for constructing superoscillatory functions is the method in \cite{Beethoven}, that constructs the signal that passes through a finite number of arbitrarily-chosen quickly oscillating amplitudes while minimizing the signal's energy, i.e., its $L^2$ norm.

This method, which is outlined in the next section, is quite versatile because by suitably choosing the prescribed points, the signals' superoscillations can be finely controlled. 
In fact, this method was used early on \cite{Ferreira2004,ferreira2002,Ferreira2006} to determine the first exact asymptotic formulas for how the minimum energy cost of superoscillatory signals scales with respect to an increase of either the frequency  or the number of the superoscillations (namely polynomially and exponentially respectively). 
Here, we will use this method to continue to study the sensitivity and scaling  properties of superoscillations. To this end, we first characterize the global shape of the function as well as the local shape of its superoscillatory stretch. In principle, when increasing the number of superoscillations, these calculations quickly become impossibly hard because increasingly ill-conditioned matrices would need to be inverted. As we will show, however, it is possible to determine a universal scaling behavior that allows one to determine significant details even of extreme superoscillatory functions that are far outside the reach of direct calculation. 

Further, we then begin to extend the scaling results for superoscillatory functions on the real line to superoscillatory functions that are periodic. We find that the behavior of periodic superoscillatory functions is remarkably similar to that of  non-periodic superoscillatory functions, albeit with some key differences. 

\section{Methods for generating superoscillations on the real line}
\label{first}
In this section, we review the construction of minimum energy superoscillations given in \cite{Beethoven} and used, e.g. in, \cite{Ferreira2004,calder,ferreira2002,Ferreira2006}. To this end, we consider the space of functions, $f$, which are bandlimited to a frequency of $\mu/2$. We can write any such $f$ in terms of its Fourier transform $\hat{f}$ as
\begin{equation}f(x)=\int_{-\mu/2}^{\mu/2} \hat{f}(\omega)e^{i\omega x}\ d\omega.\end{equation}
The aim then is to find such functions $f$ which pass through a sequence of $N$ points
\begin{equation}\{(t_0,a_0),\ (t_1,a_1),\ \cdots,\ (t_{N-1},a_{N-1})\},\end{equation}
with amplitudes of alternating sign, where the times $t_i$ are chosen sufficiently close for the resulting function to exhibit superoscillatory behavior. More specifically, the aim is to find that bandlimited function which passes through the above set of points which possesses the minimum energy. It was shown in \cite{Beethoven} that this minimum energy solution is given as a linear superposition of sinc functions centered at the interpolating points
\begin{equation}f_{\mathrm{min}}(t) = \mu\sum_{i=0}^n x_i \operatorname{sinc}\left(\mu(t-t_i)\right),\end{equation}
where the coefficients $x_i$ are given by solving the matrix equation $\mathbf{a} = \rho\mathbf{x}$, where $\mathbf{a}=(a_1,\dots, a_n)^T$ and where $\rho$ is the matrix with entries
\begin{equation}\rho_{ij} = \mu\operatorname{sinc}\left(\mu(t_i-t_j)\right),\ \ \ \ \ 0\le i,j\le N-1.
\end{equation}

For the special case of taking evenly spaced points, i.e., $t_k = k\delta$ where the spacing between consecutive points is $\delta$, the matrix $\rho$ reduces to a symmetric positive-definite matrix called the prolate matrix \cite{Slepian}.

In particular, since the prolate matrix is positive-definite, it is always invertible which is why there is a unique minimum energy solution for any choice of prescribed amplitudes in the case of uniformly spaced points. As was shown in \cite{ferreira2002,Ferreira2004}, the energy requirement for this minimum energy superoscillatory function resulting from inverting the prolate matrix scales as
\begin{equation}E[f_\mathrm{min}] = \|f_{\mathrm{min}}\|_2^2 \le \frac{\|\mathbf{a}\|^2}{\lambda^\star} \sim \|\mathbf{a}\|^2\frac{2^{4N-4}(2N-1)}{\sqrt{\pi}(\pi\mu\delta)^{2N-1}(N-1)^{3/2}},\end{equation}
where $\lambda^\star$ is the smallest eigenvalue of the prolate matrix. The inequality above attains equality if and only if the prescribed amplitudes $\mathbf{a}$ is an eigenvector of the prolate matrix with eigenvalue $\lambda^\star$.

\section{General behavior of minimum energy superoscillations}
\label{sec3}
We now investigate the general behavior of these minimum energy superoscillations. In particular, we look at their sensitivities under perturbation, and at their large and small scale behavior.

We begin with the observation that the minimum energy superoscillatory functions are determined by solving an extremely ill-conditioned matrix system: $\mathbf{a} = \rho\mathbf{c}$, where $\rho$ is the relevant prolate matrix, $\mathbf{a}$ is the vector of prescribed amplitudes, and $\mathbf{c}$ the sought-after vector of coefficients that determines the superoscillatory function as a linear combination of sinc functions.

Suppose then that $f$ is such a minimum-energy superoscillatory function fitted through $N$ points with spacing $\delta$. Let $\lambda_0 \ge \lambda_1 \ge \cdots \ge \lambda_{N-1}=\lambda^\star$ denote the $N$ eigenvalues of the relevant prolate matrix $\rho$. The key observation now is that for sufficiently rapid superoscillations, this finite sequence of eigenvalues decays very quickly, which also means that their crucial inverses obey $\lambda_{k}^{-1} \gg \lambda_{k-1}^{-1}$. In fact, the dominance grows stronger as $k$ increases. In particular, asymptotically for the prolate matrix, we have
\begin{equation}\lambda_k \sim \frac{(\delta \mu)^{2k+1}}{\delta}\frac{2^{2k}(k!)^6}{(2k+1)^2[(2k)!]^4}\prod_{j=-k}^k(N-j),\end{equation}
so that as $\delta \rightarrow 0$, we get
\begin{equation}
\frac{\lambda_{k+1}}{\lambda_{k}} \sim (\delta\mu)^2\frac{4(k+1)^6(2k-1)^2}{(2k+1)^6(2k+2)^4}[N^2-(k+1)^2].\end{equation}  \\
This shows that each eigenvalue is quadratic in $\delta$ over the previous. Therefore for sufficiently small $\delta$, each inverse eigenvalue is significantly dominant over the next. As we will now show, for many applications it is sufficient to simply consider the behavior of the smallest eigenvalue $\lambda^\star$.

For the case of evenly-spaced points, $\rho$ is a symmetric matrix and hence admits an orthonormal eigenbasis. Therefore we may write $\mathbf{a}$ in terms of the orthonormal eigenvectors of $\rho$, say $\{\mathbf{v}_k\}_{k=0}^{N-1}$, as
\begin{equation}\mathbf{a} = \sum_{i=0}^{N-1}\langle \mathbf{a},\mathbf{v}_k\rangle \mathbf{v}_k.\end{equation}
Using the fact that $\lambda_{N-1}^{-1}=(\lambda^\star)^{-1}$ is dominant, this allows us to write
\begin{equation}\mathbf{c} = \rho^{-1}\mathbf{a}= \rho^{-1}\left(\sum_{i=0}^{N-1}\langle \mathbf{a},\mathbf{v}_k\rangle \mathbf{v}_k\right) = \sum_{i=1}^N\lambda_{k}^{-1}\langle \mathbf{a},\mathbf{v}_k\rangle \mathbf{v}_k \approx \left(\lambda^\star\right)^{-1}\langle \mathbf{a},\mathbf{v}_{N-1}\rangle\mathbf{v}_{N-1}.\end{equation}

\noindent
Now we may ask three related questions regarding the shape of the superoscillations:\\

1. How sensitive is the superoscillatory stretch to errors in the prescribed coefficients. If we perturb $\mathbf{c}$ by a small amount $\Delta\mathbf{c}$, how does the perturbation propagate to the prescribed amplitudes $\mathbf{a}$?\\ 

2. What is the dependence of the overall shape of the superoscillatory function on the prescribed amplitudes? As we change the prescribed amplitudes, how wildly does the shape of the overall function vary? What is the shape of the overall function?\\ 

3. How does the superoscillatory stretch respond to prescribed amplitudes? In particular, is it possible to quantify the shape of the superoscillatory stretch?\\

\noindent The strong dominance of $1/\lambda^\star$ over the other inverse eigenvalues will allow us to address all three of these questions.

\subsection{Sensitivity to Coefficient Perturbations}
Suppose that there is some small error $\Delta \mathbf{c}$ when prescribing the coefficients of the sinc functions. Then this error propagates to $\mathbf{a}$ as $\Delta\mathbf{a}=\rho\Delta\mathbf{c}$. Now we use the fact that the first two eigenvalues of $\rho$ are much larger than the remaining, so that we can write
\begin{equation}\rho\Delta\mathbf{c} = \lambda_0\langle \Delta\mathbf{c}, \mathbf{v}_0\rangle \mathbf{v}_0 + \lambda_1\langle \Delta\mathbf{c}, \mathbf{v}_1\rangle \mathbf{v}_1.\end{equation}
The vectors $\mathbf{v}_{k}$ are known as the Discrete Prolate Spheroidal Sequences (DPSS). It is an even vector (in the sense that the $i$th entry of the vector is equal to the $(N-i)$th entry) when $k$ is even, and an odd vector (in the sense that the $i$th entry of the vector is equal to the negative of the $(N-i)$th entry) when $k$ is odd \cite{Slepian}. 

This shows that the general shape of the perturbation is of the form of a vertical displacement and stretch (due to $\mathbf{v}_0$) and that of a tilt (due to $\mathbf{v}_1$). The relative magnitude of the perturbation is then
\begin{equation} \frac{\|\Delta\mathbf{a}\|}{\|\mathbf{a}\|} = \frac{\|\rho\Delta\mathbf{c}\|}{\|\mathbf{a}\|} \approx \frac{1}{\|\mathbf{a}\|}\sqrt{\lambda_0^2\langle \Delta\mathbf{c}, \mathbf{v}_0\rangle^2 + \lambda_1^2\langle \Delta\mathbf{c}, \mathbf{v}_1\rangle^2}.\end{equation}
How small must we make $\Delta\mathbf{c}/\mathbf{c}$ so that the expression above will be much less than $1$? We can write the above as
\begin{eqnarray} \frac{1}{\|\mathbf{a}\|}\sqrt{\lambda_0^2\langle \Delta\mathbf{c}, \mathbf{v}_0\rangle^2 + \lambda_1^2\langle \Delta\mathbf{c}, \mathbf{v}_1\rangle^2}\nonumber\\ =\left(\lambda^\star\right)^{-1}\left\langle \frac{\mathbf{a}}{\|\mathbf{a}\|},\mathbf{v}_{N-1}\right\rangle\frac{\|\Delta\mathbf{c}\|}{\|\mathbf{c}\|} \sqrt{\lambda_0^2\left\langle \frac{\Delta\mathbf{c}}{\|\Delta\mathbf{c}\|},\mathbf{v}_0\right\rangle^2 + \lambda_1^2\left\langle \frac{\Delta\mathbf{c}}{\|\Delta\mathbf{c}\|}, \mathbf{v}_1\right\rangle^2}.
\end{eqnarray}

\noindent
The typical values of the terms 
$$\left\langle \frac{\mathbf{a}}{\|\mathbf{a}\|},\mathbf{v}_{N-1}\right\rangle,\ \ \ \text{and}\ \ \ \sqrt{\lambda_0^2\left\langle \frac{\Delta\mathbf{c}}{\|\Delta\mathbf{c}\|},\mathbf{v}_0\right\rangle^2 + \lambda_1^2\left\langle \frac{\Delta\mathbf{c}}{\|\Delta\mathbf{c}\|}, \mathbf{v}_1\right\rangle^2}$$ are completely negligible in the above equation as compared to $\left(\lambda^{\star}\right)^{-1}$, therefore for $\frac{\|\Delta\mathbf{a}\|}{\|\mathbf{a}\|} \ll 1$, we must have
\begin{equation} \frac{\|\Delta\mathbf{c}\|}{\|\mathbf{c}\|}\left(\lambda^\star\right)^{-1} \ll 1.\end{equation}
Note that this is an extremely stringent requirement on the sensitivity of the coefficients. In particular, recall that the energy of the function scales as $E[f] \approx \left(\lambda^\star\right)^{-1}$, so that we will require a sensitivity on the order of
\begin{equation}\frac{\|\Delta\mathbf{c}\|}{\|\mathbf{c}\|} \ll \frac{1}{E[f]} = \frac{1}{\|f\|_2^2}.\end{equation}
This condition means that robust superoscillations will be extremely difficult to construct in the lab and that any existing superoscillations will be extremely susceptible to any form of dispersion in its medium of propagation. On the other hand, anytime a phenomenon exhibits extreme sensitivity, there is a chance that it will be useful for metrology, see also \cite{prain}. We also note in passing that superoscillations can be made more robust numerically by considering non-centered sinc functions at the cost of a little energy and amplitude \cite{LeeFerreira}. 

\subsection{Large-Scale Shape of the Superoscillatory Functions}
In this section, we answer the second question: Consider the case of prescribing $N$ equally spaced points with spacing $\delta$. How does the overall shape of the superoscillatory function change as we vary the prescribed amplitudes, $\mathbf{a}$?

Consider the normalized eigenvector $\mathbf{v}_{N-1}$ to the smallest eigenvalue $\lambda^\star$ of $\rho$. If we take the prescribed amplitudes to be this eigenvector, i.e., $\mathbf{a} = \mathbf{v}_{N-1}$, then we get a particular minimum energy superoscillatory function $\tilde{f}$ passing through the amplitudes $\mathbf{v}_{N-1}$. The function $\tilde{f}$ is the function of largest energy out of all such minimum energy superoscillatory functions prescribed with unit norm amplitudes. The vector $\mathbf{v}_{N-1}$ will be either even or odd depending on whether $N-1$ is even or odd \cite{Slepian}. This means that the resulting function $\tilde{f}$ is either an even function (for an odd number of prescribed points) or an odd function (for an even number of prescribed points) about the center of its superoscillatory stretch.

\begin{figure}[ht]
    \centering
    \includegraphics[width=0.85\textwidth]{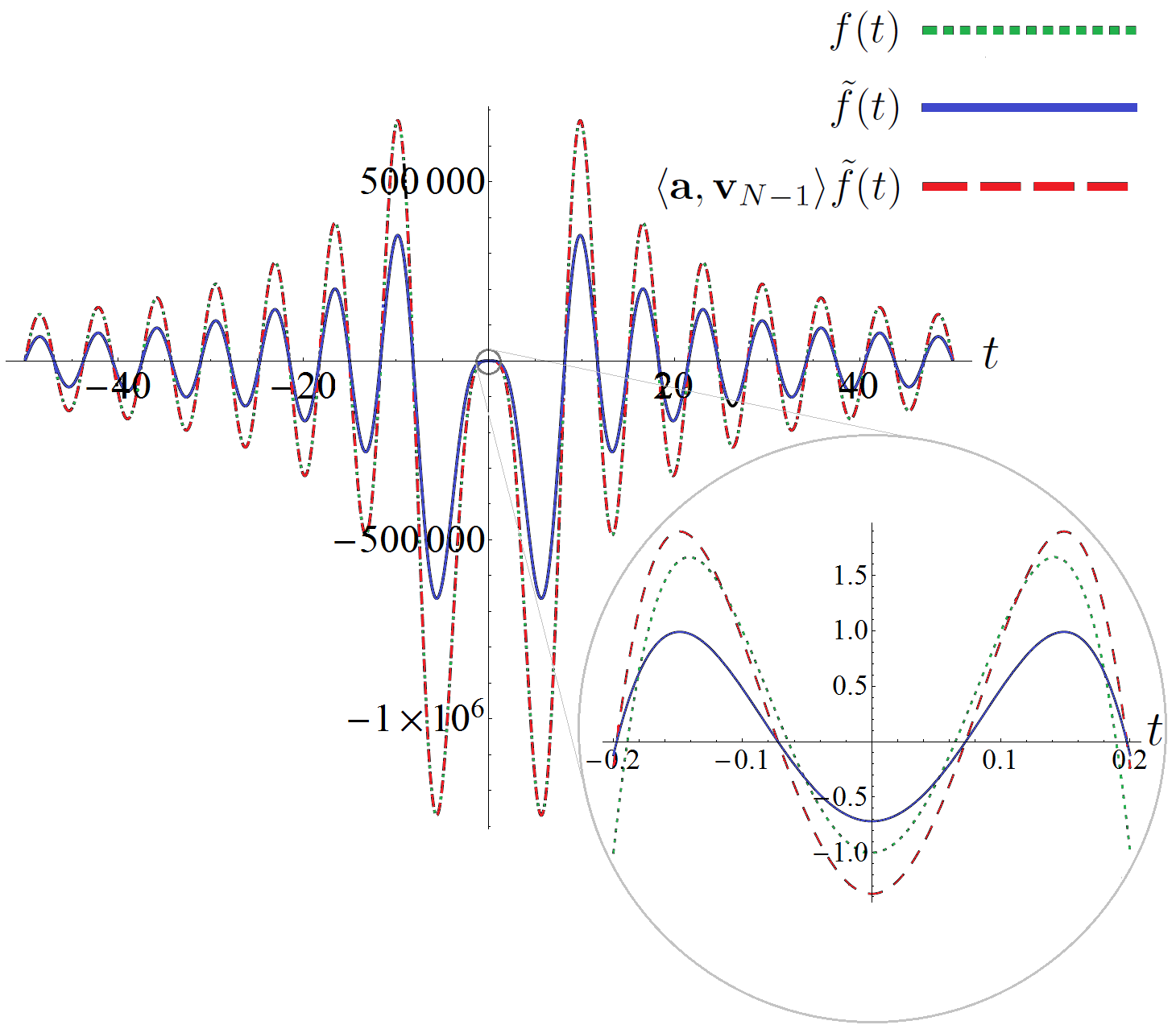}
    \caption{Comparison of $f$ versus $\tilde f$ (both scaled and unscaled). Superoscillatory stretch of the functions shown in zoom.}
    \label{fig:compare}
\end{figure}

In Figure \ref{fig:compare}, $f$ is a superoscillatory function with bandwidth $\mu=1$ fitted through the $5$ points 
$$\left\{(-1/5,-1),\ (-1/10,1),\ (0,-1),\ (1/5,1),\ (1/5,-1)\right\}.$$ 
It is plotted against the corresponding energetically most expensive function $\tilde{f}$, both unscaled and scaled by scalar multiple $\langle\mathbf{a},\mathbf{v}_{N-1}\rangle\tilde f$. The superoscillatory stretch of the functions are shown in the zoom and can clearly be seen to be distinct. However, note that in the original image, the functions $f$ and $\langle\mathbf{a},\mathbf{v}_{N-1}\rangle\tilde f$ overlap completely as indicated by the dotted-dashed curve.

The shape of $\tilde{f}$ is highly characteristic of general superoscillatory functions for generic prescribed amplitudes. Indeed, a generic superoscillatory function will essentially be a scalar multiple of $\tilde{f}$ in the sense that the relative difference between the two functions is very small. Consider a general superoscillatory function $f$ constructed from a linear combination of centered sincs, which we will denote as $f_i$. Then we may write
\begin{equation} f(x) = \sum_{i=0}^N c_i f_i(x) = \mathbf{c}\cdot \mathbf{f}(x), \end{equation}
where in the last step we've collected the functions $f_i$ as the vector $\mathbf{f}(x) = (f_0(x),\ f_1(x),\ \cdots,\ f_n(x))^\mathrm{T}$. The coefficients $c_i$ are given by $\mathbf{c} = \rho^{-1}\mathbf{a}$ and the vector $\mathbf{c}$ is heavily dominated by the leading term in its eigenvector expansion. Therefore we can write the relative difference between the two vectors as
\begin{equation}\frac{\|\mathbf{c}-\left(\lambda^\star\right)^{-1}\langle \mathbf{a},\mathbf{v}_{N-1}\rangle \mathbf{v}_{N-1}\|}{\|\mathbf{c}\|} \ll 1.\end{equation}
Then we have
\begin{equation}\left| \langle\mathbf{a},\mathbf{v}_{N-1}\rangle\tilde{f}(x) - f(x)\right| = \left|\left(\left(\lambda^\star\right)^{-1}\langle\mathbf{a},\mathbf{v}_{N-1}\rangle\mathbf{v}_{N-1} - \mathbf{c}\right)\cdot\mathbf{f}(x)\right| \ll \|\mathbf{c}\|\|\mathbf{f}(x)\| = \frac{|f(x)|}{\cos \theta},\end{equation}
where $\theta$ is the angle between the vectors $\mathbf{c}$ and $\mathbf{f}(x)$. Note that $\cos \theta = 0$ if and only if $f(x) = 0$ so that the relative difference can be large near the zeros of the function, and in fact it is divergent there. But this divergence is superficial in the sense that it arises simply because the zeros of the functions do not coincide exactly. Away from the zeros of the function, the term $\cos \theta$ is roughly of order unity so that $\langle \mathbf{a},\mathbf{v}_{N-1}\rangle \tilde{f}$ closely approximates $f$ in terms of relative difference,
\begin{equation}|\langle \mathbf{a},\mathbf{v}_{N-1}\rangle\tilde{f} - f| \ll |f|.\end{equation} 
This result is sufficient to fix the overall shape of the function. Since the function $f$ is generic in the above arguments, the overall shape of any generic superoscillatory function is well approximated by the single function representative function $\tilde{f}$, up to a scalar multiple. Consequently, the overall shape of a generic superoscillatory function is extremely stable under changes to the prescribed amplitudes. This result curiously contrasts with the fact that the superoscillatory stretch is extremely sensitive to coefficient perturbations.

\subsection{The Shape of the Superoscillatory Stretch}

In this section we address the third question posed above: what are general properties of the superoscillatory stretch?

To this end, let $f$ be a minimum energy superoscillatory function through $N$ points on some superoscillatory interval $I$.\footnote{While we mostly consider equally spaced points in this paper, the results of this section are not limited to equally spaced points.} We found strong numerical evidence which suggests that the restriction of $f$ to $I$ is very closely approximated by the unique interpolating polynomial of least degree through the points prescribed for $f$. The reason for the polynomial behavior of the superoscillatory stretch is not well understood. While there lacks analytical proof for the emergence of this behavior, there is sufficient numerical evidence for us to make the following conjecture:\\

\noindent
\bf Conjecture: \rm Let $f$ be a minimum energy superoscillatory function fitted through $N$ (not necessarily equally-spaced) points with $x$-coordinates $x_1 < x_2 < \cdots < x_N$. Let $\delta = \max_{1\le i \le N-1}\left|x_{i+1}-x_i\right|$ and let $p(x)$ denote the unique polynomial of least degree passing through the same $N$ points as $f$. Then for any $\epsilon > 0$, there exists $\delta > 0$ such that 
$$\| f(x)-p(x)\|_I < \epsilon,$$
where $\|\cdot \|_I$ is the sup-norm restricted to the interval $I=[x_1,x_N]$.\\

That is to say, the absolute difference between the interpolating polynomial and the superoscillatory function $f$ is vanishing as the function becomes increasingly superoscillatory. More precisely, numerical evidence suggests that the absolute difference between $f$ and $p$ scales roughly as $\delta^2$. In fact the approximation appears to be remarkably accurate, with small errors as soon as the function enters the superoscillatory regime (i.e., when $\delta$ becomes less than the Nyquist spacing). The interpolating polynomial is a much better approximation to the function than the Taylor polynomial of the same (or even slightly higher) degree, especially in the regime where $f$ is not very superoscillatory, i.e., the regime of large $\delta$.

\begin{figure}[ht]
    \centering
    \includegraphics[width=0.85\textwidth]{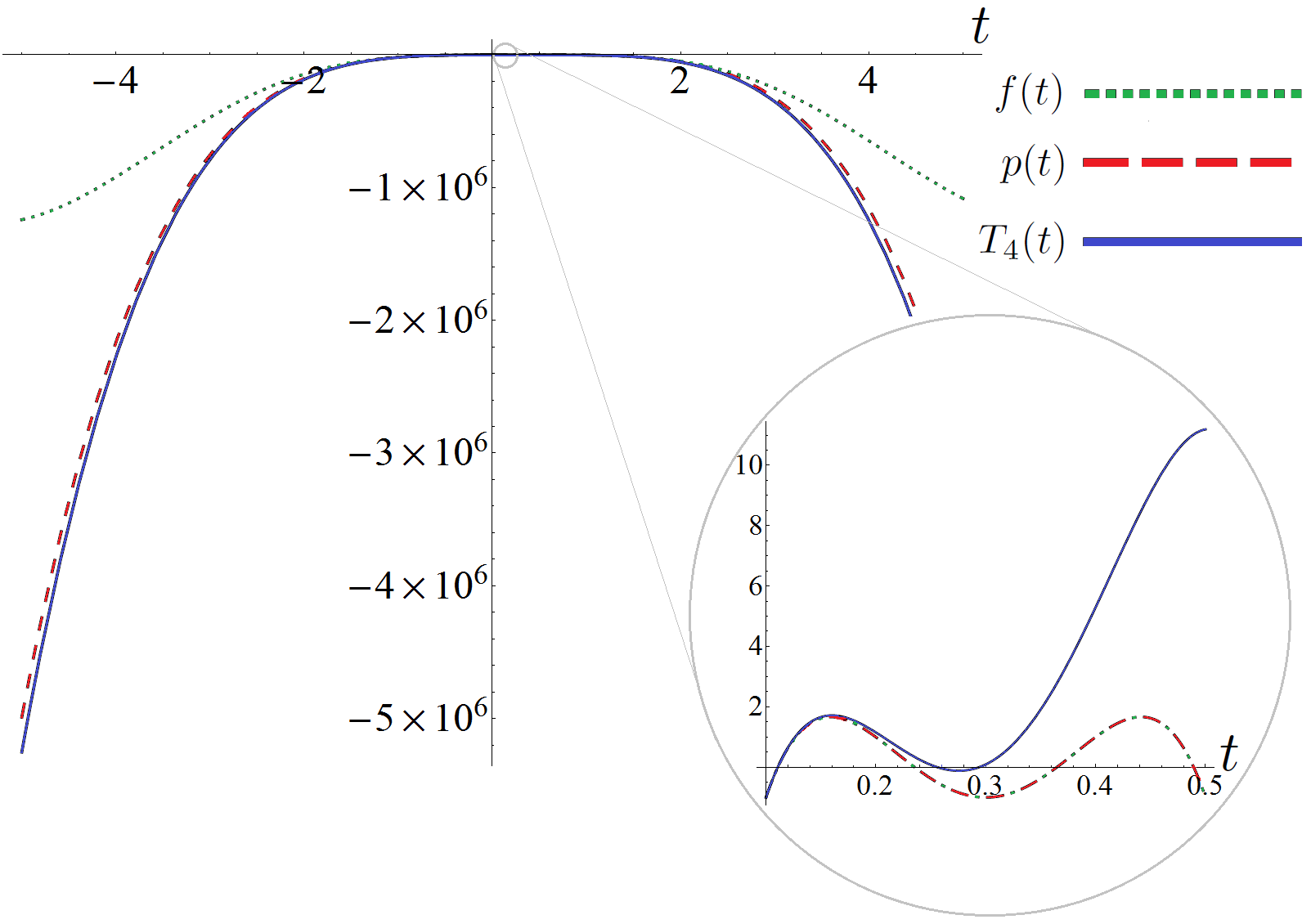}
    \caption{Comparison of $f$ versus least degree approximating polynomial $p$ and $4$th degree Taylor polynomial $T_4$. Superoscillatory stretch of the functions shown in zoom.}
    \label{fig:poly}
\end{figure}

In Figure \ref{fig:poly}, a superoscillatory function $f$ through the points 
$$\{(1/10,-1),\ (1/5,1),\ (3/10,-1),\ (2/5,1),\ (1/2,-1)\}$$ 
is plotted against the least degree interpolating polynomial (of degree $4$) through the same points. Also plotted is $T_4(t)$, the $4$th degree Taylor polynomial of $f(t)$. In the zoom of the superoscillatory stretch, the polynomial $p$ and the superoscillatory function $f$ overlap and are indistinguishable in the plot, as indicated by the dotted-dashed curve. In contrast, the $4$th degree Taylor polynomial is clearly seen as a poorer approximation to $f$ than $p$.

We also remark that there has been recent work showing that superoscillatory functions can be made to uniformly approximate an arbitrary polynomial, with arbitrary accuracy, on an interval \cite{Chremmos}. 


The polynomial behavior of the superoscillatory stretch helps to explain the characteristic shape of the function on the stretch. In particular, note that if the number of equidistantly prescribed points is increased, the superoscillatory stretch tends to exhibit oscillations of increasing amplitude towards the ends of the superoscillatory interval. This happens even when the prescribed points are not oscillatory themselves. We can now interpret this as an example of Runge's phenomenon \cite{apprx}.

Furthermore, knowing that the behavior of the superoscillatory stretch is polynomial is of tremendous advantage because knowing the interpolating method suggests modifications to the interpolating points. For example, if we wish to mitigate Runge's phenomena we would prescribe points according to Chebyshev nodes. Likewise, if we wish to find the closest approximation to a given continuous function in $L^\infty$ norm, then we prescribe points according to the appropriate minimax approximating polynomial. In simple terms, knowing that the superoscillatory stretch behaves roughly as a polynomial allows us to subsume the study of superoscillations under the study of the appropriate polynomials.

\subsubsection{Fourier transform of the Superoscillatory stretch.}

Under the conjecture of the previous section, we can also draw conclusions about the Fourier transform of the superoscillatory stretch. Since the superoscillatory stretch is well-approximated by a polynomial, for sufficiently superoscillatory functions we may write
\begin{equation}\|f(x) - p(x)\|_2 < \sqrt{|I|}\|f(x)-p(x)\|_I<\epsilon,\end{equation}
where $\|\cdot\|_2$ is the $L^2$ norm on the superoscillatory interval $I$, $\|\cdot\|_I$ is the sup-norm on the superoscillatory interval $I$, $p(x)$ the approximating polynomial, and where $\epsilon$ can be made arbitrarily small with sufficiently superoscillatory functions. Since the Fourier transform is unitary, i.e. it preserves the $L^2$ norm, it follows that the Fourier transform of a sufficiently rapidly superoscillating function can be reduced to simply studying the Fourier transforms of polynomials truncated on an interval.

Specifically, without loss of generality, we can study the truncated Fourier transform of the monomials on the interval $[-L,L]$. These are of the form, for even and odd monomials respectively,
\begin{equation}\int_{-L}^L x^{2n}e^{-itx}\ dx = \frac{2(A_{n}(Lt)\sin(Lt) + B_{n}\cos(Lt))}{t^{2n+1}},\end{equation}
\begin{equation}\int_{-L}^L x^{2n+1}e^{-itx}\ dx = \frac{2i(C_{n}(Lt)\sin(Lt) + D_{n}\cos(Lt))}{t^{2n+2}},\end{equation}
where $A,B,C,D$ are polynomials satisfying the recursive relations
\begin{equation}A_{n}(t) = t^{2n}-2n(2n-1)A_{n-1}(t),\ \ \ \ A_0(t)=1,\end{equation}
\begin{equation}B_{n}(t) = (2n)t^{2n-1}-2n(2n-1)B_{n-1}(t),\ \ \ \ B_0(t)=0,\end{equation}
\begin{equation}C_{n}(t) = -(2n+1)t^{2n}-(2n+1)(2n)C_{n-1}(t),\ \ \ \ C_0(t)=-1,\end{equation}
\begin{equation}D_{n}(t) = t^{2n+1}-(2n+1)(2n)D_{n-1}(t),\ \ \ \ D_0(t)=t.\end{equation}
All of the above relations can be easily proven by repeated integration by parts. Together, these relations serve to give a simple analytic approximation to the Fourier transforms of extreme superoscillatory functions and can be useful for analyzing the spectrum of the superoscillating stretch.

\subsubsection{Consequences for Extreme Superoscillatory Functions.}
The results of previous sections suggest that the minimum energy superoscillatory functions are in one sense fragile, but also in another sense very rigid. Their global shape is essentially determined up to multiplicative factor by the function $\tilde{f}$ fitted through the eigenvector to the smallest eigenvalue of the Prolate matrix. Now we see that the local behavior of the superoscillatory stretch is essentially polynomial and that the function is well-approximated by the interpolating polynomial of least degree.

Most importantly, note that superoscillatory functions are difficult to construct. Their sensitivity to perturbations suggests that even the numerical study of extremely superoscillatory functions is very hard. Therefore, being able to obtain a good approximation of the behavior of an extremely superoscillatory function is of great interest. Indeed, the results of section \ref{sec3} offer a rather complete picture of what an extreme superoscillatory function looks like. Consider a minimum energy superoscillatory function $f$ fitted through $N$ points. We can piece together the rough behavior of $f$ as follows: The large scale behavior of the function outside the superoscillatory segment is, up to a scalar multiple, provided by the least energy solution $\tilde f$ fitted through the eigenvector corresponding to smallest eigenvalue of the prolate matrix. Further, the fine structure of the superoscillatory stretch is provided by the interpolating polynomial of least degree fitted through the same points that determine $f$.

\section{Periodic Superoscillations}

Let us now investigate to what extent the properties of superoscillatory functions on the real line carry forward to superoscillations in periodic functions. Periodic superoscillatory functions are of interest for practical applications, for example when designing wave forms by superimposing laser beams or any other sources of monochromatic radiation. 

Therefore, our aim now is to generalize the previous results to construct also minimum energy superoscillations with a specific periodicity. The mathematical framework now changes in that sinc functions are not available in a space of bandlimited functions that are periodic since they possess unlimited support.   To this end, let $f$ be a function with period $2\pi$. Recall that $f$ can be expressed as a Fourier series given by
\begin{equation}f(t)=\sum_{n=-\infty}^\infty c_ne^{int}.\end{equation}
Consider the case of $f$ being bandlimited to a frequency of $M$, by which we mean that the highest component of the Fourier series is $e^{\pm iMt}$:
\begin{equation}f(t) = \sum_{n=-M}^M c_n e^{int}.\end{equation}
As before, we prescribe points to create superoscillations. We require our function $f$ to pass through the $N$ points
\begin{equation}\{(t_0,a_0),\ (t_1,a_1),\ \cdots,\ (t_{N-1},a_{N-1})\},\end{equation}
by which we obtain a resulting matrix equation of the form
\begin{equation}\begin{pmatrix} a_0 \\ a_1 \\ \vdots \\ a_{N-1}\end{pmatrix} = \begin{pmatrix}1 & e^{it_0} & e^{-it_0} & e^{2it_0} & e^{-2it_0} & \cdots & e^{Mit_0} & e^{-Mit_0}\\
1 & e^{it_1} & e^{-it_1} & e^{2it_1} & e^{-2it_1} & \cdots & e^{Mit_1} & e^{-Mit_1}\\
\vdots & \vdots & \vdots & \vdots & \vdots & \ddots & \vdots & \vdots \\
1 & e^{it_{N-1}} & e^{-it_{N-1}} & e^{2it_{N-1}} & e^{-2it_{N-1}} & \cdots & e^{Mit_{N-1}} & e^{-Mit_{N-1}}\end{pmatrix}
\begin{pmatrix}c_0 \\ c_1 \\ c_{-1} \\ \vdots \\ c_M \\ c_{-M}\end{pmatrix},\end{equation}
which we will abbreviate as $\mathbf{a} = \mathrm{T}\mathbf{c}$, where $\mathbf{a}$ is the vector of amplitudes, $\mathbf{c}$ the vector of (complex) coefficients and $T$ the $N\times(2M+1)$ matrix of complex exponentials. By Parseval's theorem, the norm of a periodic function with Fourier coefficients $c_n$ is given by
\begin{equation} \|f\|^2 = \frac{1}{2\pi}\int_{-\pi}^\pi |f(t)|^2\ dt = \sum_{n=-\infty}^\infty |c_n|^2,\end{equation}
and for $f$ bandlimited to $M$, this is precisely the norm of the coefficient vector:
\begin{equation} \|f\|^2  = \|\mathbf{c}\|^2 = \mathbf{c}^\dagger\mathbf{c},\end{equation}
where $\dagger$ denotes the Hermitian transpose. Therefore to find the minimum energy solution is to find the vector of coefficients with the minimum norm. The minimum energy solution can be constructed from the Moore-Penrose pseudo-inverse through
\begin{equation}\mathbf{c}_\mathrm{min} = \mathrm{T}^+\mathbf{a},\end{equation}
where $T^+$ denotes the pseudo-inverse of $\mathrm{T}$.\\

\noindent 
\bf Proposition: \rm For distinct values of $t_k$, the matrix $T$ has full rank.\\

\noindent
\bf Proof: \rm Rearrange the columns of $T$ and factor out $e^{-iMt_k}$ from row $k+1$. Call the resulting matrix $T'$ and note that $T$ and $T'$ have the same rank. Now, the rows of $T'$ are geometric sequences 
\begin{equation}T' = \begin{pmatrix}1 & e^{it_0} & e^{2it_0} & e^{3it_0} & \cdots & e^{2Mit_0}\\
1 & e^{it_1} & e^{2it_1} & e^{3it_1} & \cdots & e^{2Mit_1}\\
\vdots & \vdots & \vdots & \vdots & \ddots & \vdots \\
1 & e^{it_{N-1}} & e^{2it_{N-1}} & e^{3it_{N-1}} & \cdots & e^{2Mit_{N-1}}\end{pmatrix}.\end{equation}
If $2M+1 \ge N$, then taking the first $N$ columns, we find a $N\times N$ Vandermonde sub-matrix. For distinct values of $t_k$, the Vandermonde matrix is invertible, so we have a sub-matrix of rank $N$. Since $T'$ has a rank $N$ sub-matrix, it follows that it has full row-rank $N$. Likewise, if $2M+1 \le N$, then repeat the argument with the $(2M+1)\times (2M+1)$ Vandermonde sub-matrix obtained from the first $2M+1$ rows. Again, we find that the matrix has full column-rank $2M+1$. $\square$\\

In the case that $N\le 2M+1$, the matrix $T$ has full row-rank, and so $\mathrm{T}\mathrm{T}^\dagger$ is invertible and the pseudo-inverse reduces to
\begin{equation}\mathrm{T}^+ = \mathrm{T}^\dagger\left(\mathrm{T}\mathrm{T}^\dagger\right)^{-1}.\end{equation}
For $N > 2M+1$ however, the matrix $TT^\dagger$ will not be invertible. This corresponds to the fact that it is not possible to prescribe more points than the dimension of our function space ($2M+1$).

Let $\mathrm{S} = \mathrm{T}\mathrm{T}^\dagger$. Then $\mathrm{S}$ is a real-valued, symmetric, Toeplitz matrix with entries given by
\begin{equation}\mathrm{S}_{jk} = D_M\left(t_j-t_k\right),\ \ \ \ \ 0\le j,k\le N-1,\end{equation}
where $D_M(t)$ denotes the Dirichlet kernel, given by
\begin{equation}D_M(t) = \frac{\sin\left((M+\frac{1}{2})t\right)}{\sin\left(\frac{t}{2}\right)} = \sum_{n=-M}^M e^{int}.\end{equation}

In various ways, the Dirichlet kernels play the counter-parts of the sinc functions for the case of periodic functions. Indeed, we shall see that the minimum energy periodic superoscillatory solutions are precisely obtained as a linear superposition of Dirichlet kernel functions, similar to how real line minimum energy superoscillations are obtained as a linear superposition of sinc functions.\\

\noindent 
\bf Proposition\rm: Let
\begin{equation}\{(t_0,a_0),\ (t_1,a_1),\ \cdots,\ (t_{N-1},a_{N-1})\}\end{equation}
denote a sequence of $N$ points. Then for each integer $M$ with $N \le 2M+1$, we have a minimum energy periodic function bandlimited to $M$, passing through the above prescribed points. Moreover, the minimum energy solution is given by
\begin{equation}f_\mathrm{min}(t) = \sum_{k=0}^{N-1}x_k D_M(t-t_k),\end{equation}
where the coefficients $\{x_k\}_{k=0}^{N-1}$ are uniquely determined by the prescribed amplitudes $\{a_k\}_{k=0}^{N-1}$. The energy of the function satisfies
\begin{equation}\|f\|^2 \le \frac{\|\mathbf{a}\|^2}{\lambda^\star_\mathrm{per}},\end{equation}
where $\lambda^\star_\mathrm{per}$ is the smallest eigenvalue of the matrix $\mathrm{S}$.\\

\noindent
\bf Proof\rm: From before, for $N \le 2M+1$, the matrix $\mathrm{S}$ has full-rank (as long as the values of $\{t_k\}$ are pairwise unequal). This means that $\mathrm{S}\mathbf{c} = \mathbf{a}$ is satisfiable for any choice of $\mathbf{a}$. The minimum norm solution is given by
\begin{equation}\mathbf{c}_\mathrm{min}=T^+\mathbf{a} = \mathrm{T}^\dagger\left(\mathrm{T}\mathrm{T}^\dagger\right)^{-1}\mathbf{a}.\end{equation}
The function corresponding to the minimum norm solution is
\begin{equation}g(t) = \sum_{n=-M}^M c_ne^{int},\end{equation}
where the coefficients $\{c_n\}$ are the entries of $\mathbf{c}_\mathrm{min}$. Let us show that this can be put into the form in equation $(*)$. Let us define $\mathbf{x} = \mathrm{S}^{-1}\mathbf{a}$, and note that the entries of $\mathbf{x}$ are precisely the coefficients required in equation $(17)$ for $f(t)$ to pass through the prescribed points. Note that we have
\begin{equation}\mathbf{c}_{\mathrm{min}} = \mathrm{T}^\dagger\mathrm{S}^{-1}\mathbf{x} = \mathrm{T}^\dagger \mathbf{x}.\end{equation}

\noindent Let us now show that $f=g$. Indeed, we get
\begin{equation}f(t) = \sum_{k=0}^{N-1}x_k D_M(t-t_k) = \sum_{k=0}^{N-1}x_k\sum_{n=-M}^Me^{in(t-t_k)} = \sum_{n=-M}^Me^{int}\left[\sum_{k=0}^{N-1}x_ke^{-int_k}\right].\end{equation}
Comparing the above with the form of $g$, we have $f=g$ if and only if
\begin{equation}\sum_{k=0}^{N-1}x_ke^{-int_k} = c_n\end{equation}
for $n=0,1,\cdots,N-1$. But this system of equations exactly corresponds to the fact that
\begin{equation}\mathbf{c}_\mathrm{min} = T^\dagger\mathbf{x},\end{equation}
which we had previously established. This shows that $f=g$ so that the minimum energy solution takes the form of $(17)$ with coefficients $\mathbf{x}$ determined by $\mathbf{x} = \mathrm{S}^{-1}\mathbf{a}$. Now the norm of the solution is given by
\begin{equation}\|f\|^2=\mathbf{c}_\mathrm{min}^\dagger\mathbf{c}_\mathrm{min} = \mathbf{a}^\dagger \mathrm{S}^{-1}\mathrm{T}\mathrm{T}^\dagger \mathrm{S}^{-1}\mathbf{a}= \mathbf{a}^\dagger S^{-1}\mathbf{a}.\end{equation}
From the above expression, we see that the norm is maximized precisely when $\mathbf{a}$ is an eigenvector to $\mathrm{S}^{-1}$ of the largest eigenvalue, or equivalently, an eigenvector of $\mathrm{S}$ with the smallest eigenvalue $\lambda^\star_\mathrm{per}$. Therefore we have
\begin{equation}\|f\|^2 \le \frac{\|\mathbf{a}\|^2}{\lambda^\star_\mathrm{per}},\end{equation}
with equality if and only if $\mathbf{a}$ is an eigenvector to $\mathrm{S}$ with smallest eigenvalue $\lambda^\star_\mathrm{per}$. $\square$\\

Note that the minimum energy solution will be real-valued if the prescribed amplitudes $\mathbf{a}$ are real. This is because the coefficients $\mathbf{x}$ are determined by $\mathbf{x} = \mathrm{S}^{-1}\mathbf{a}$ and $S$ is a real-valued matrix.

\begin{figure}[ht]
    \centering
    \includegraphics[width=0.85\textwidth]{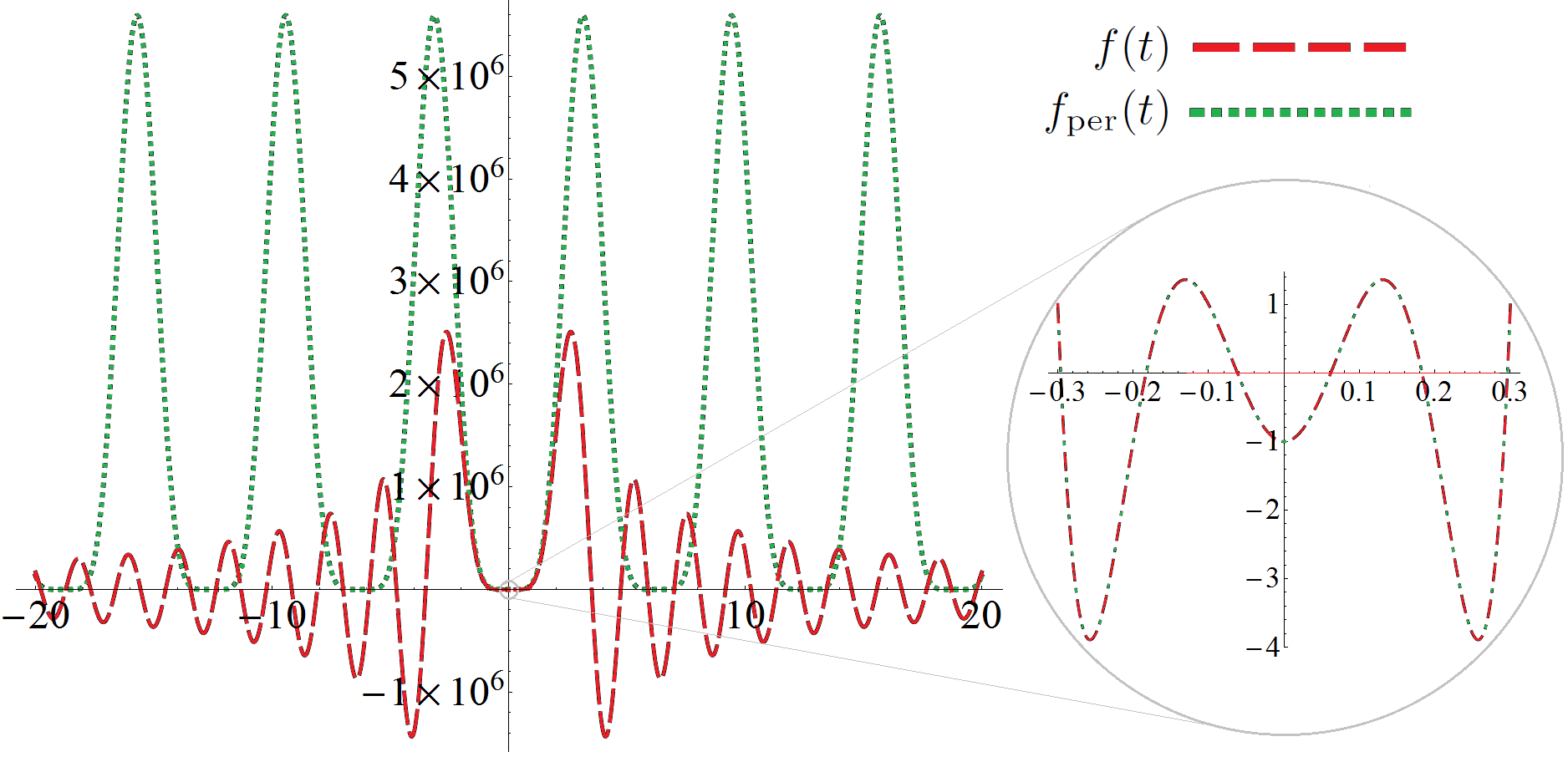}
    \caption{Comparison of real line $f$ versus periodic $f_\mathrm{per}$. Superoscillatory stretch of the functions shown in zoom.}
    \label{fig:periodic}
\end{figure}

Figure \ref{fig:periodic} shows a periodic minimum energy superoscillatory function $f_\mathrm{per}$ with $M=3$ fitted through the points 
$$\{(-3/10,1),\ (-1/5,-1),\ (-1/10,1),\ (0,-1),\ (1/10,1),\ (1/5,-1),\ (3/10,1)\}.$$ 
The resulting function is $2\pi$-periodic. The corresponding non-periodic minimum energy superoscillatory $f$ is plotted alongside $f_\mathrm{per}$. First, note that the superoscillatory portions of the two functions agree well and are virtually indistinguishable in the plot, as indicated by the dotted-dashed line. Since it is expected that the smallest eigenvalue of the matrix $S$ for periodic superoscillations is also very dominant, the results of the previous section should continue to hold for periodic superoscillations as well. In particular, we expect the overall shape of the periodic superocillations to be given by a particular function $\tilde{f}_\mathrm{per}$ much like the case for real line superoscillations. We also expect the superoscillatory stretch of the periodic superoscillations to be well approximated by a respective polynomial. 

Secondly, note that the peak of the periodic function appears much larger than the peak of the non-periodic function. It appears in general that the $L^2$-norm of $f_\mathrm{per}$ over a single period is comparable to the $L^2$-norm of $f$ over the entire real line (as we will show later), so there is a non-trivial amplitude cost associated with forcing a superoscillating signal to be periodic.

The matrix $S$ defined in this section plays the exact same role as the prolate matrix $\rho$ for superoscillations on the real line. In particular, it is because we have precise asymptotics for the prolate matrix that we know the energy growth behavior for superoscillations on the real line. Studying the eigenvalues of $S$ will likewise lead to the growth behavior of periodic superoscillations.

\begin{figure}[ht]
    \centering
    \includegraphics[width=1\textwidth]{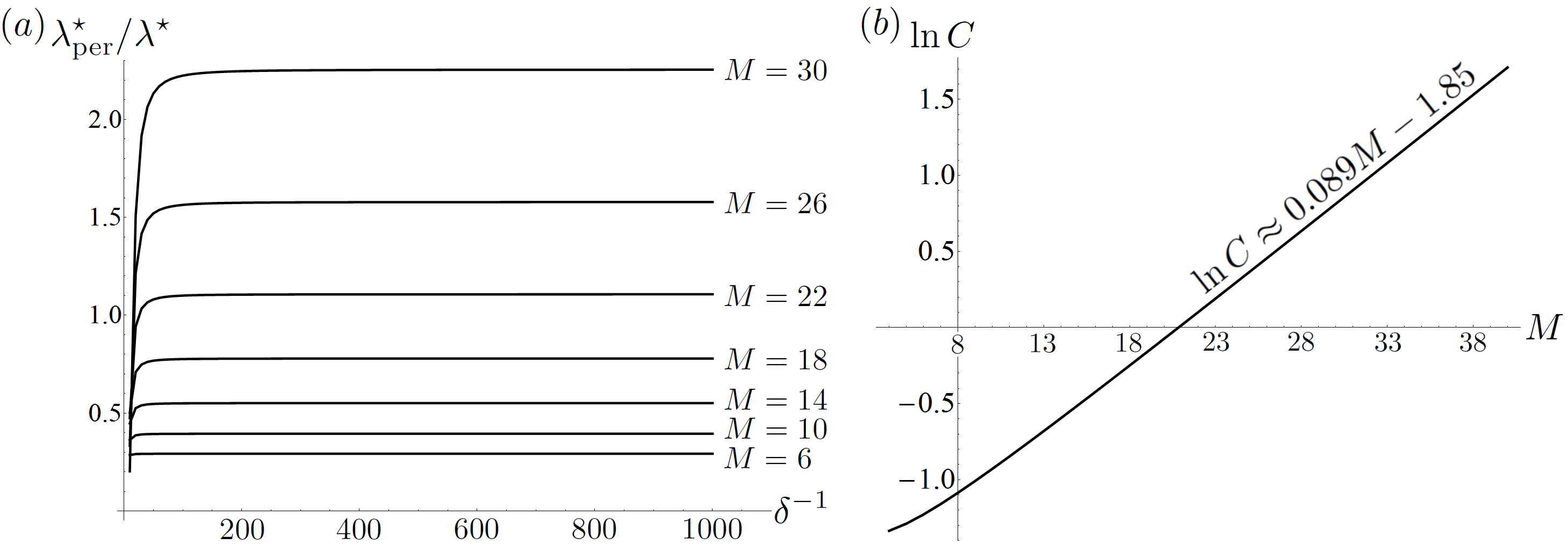}
    \caption{a) Comparison of $\lambda^\star_\mathrm{per}$ with $\lambda^\star$ as a function of $\delta^{-1}$, for various values of $M$.\ \ b) Plot of $\ln C$ as a function of $M$. The equation for the least squares fit is also shown.}
    \label{fig:num1}
\end{figure}

Let us examine the energy behavior of the periodic superoscillations numerically. Note that while $S$ and the prolate matrix appear to be very similar, there are key differences which complicates the analysis. In particular, one key difference is that for a fixed bandlimit, we may prescribe as many points as desired in the case of real line superoscillations; the resulting prolate matrix will always be non-singular. The same cannot be said for the matrix $S$, as we are bound by the requirement $N \le 2M+1$ for $S$ to be non-singular. With these considerations in mind, we will pick a simple case for analysis in which the bandlimit increases together with number of prescribed points. Explicitly, let us consider the case of a $2\pi$-periodic superoscillation with bandlimit $M$, fitted through $M$ equally spaced points with spacing $\delta$. Then $S$ is a $M \times M$ symmetric square matrix with entries given by
\begin{equation}S_{ij}=D_M(\delta(i-j)).\end{equation}
For comparison, we will consider a real line superoscillation fitted through the same $M$ points, with a bandlimit of $\mu = M$. This corresponds to studying the $M\times M$ prolate matrix $\rho$ with entries given by
\begin{equation}\rho_{ij} = M\operatorname{sinc}(M\delta(i-j)).\end{equation}
To examine the asymptotic behavior of the superoscillations, it is sufficient to examine the smallest eigenvalues of $S$ (call it $\lambda^\star_\mathrm{per}$) and $\rho$ (call it $\lambda^\star$). It turns out that there is a close correspondence between the asymptotic behavior of the two matrices, as shown in Figure \ref{fig:num1}a. 

From Figure \ref{fig:num1}a, it appears that the ratio $\lambda^\star_{\mathrm{per}}/\lambda^\star$ quickly tends to a definite limit (dependent on $M$) as $\delta \rightarrow 0$. For definiteness, let's call the limit of this ratio $C(M)$, so that we can write
\begin{equation}\lim_{\delta \rightarrow 0}\frac{\lambda^\star_\mathrm{per}}{\lambda^\star} = C(M).\end{equation}
This suggests that for fixed $M$, we have $\lambda^\star_\mathrm{per} \sim c(M)\lambda^\star$. We can also examine the behavior of $C$ as a function of $M$, Figure \ref{fig:num1}b shows the behavior of $\ln C$ as a function of $M$, which we see to be asymptotically linear. The least square fit of the data suggests that $\ln C$ is well approximated as
\begin{equation}\ln C \approx 0.089M - 1.85\ \ \ \ \text{or}\ \ \ \ \ C(M) \approx 0.157\cdot(1.093)^M.\end{equation}
This suggests that the full asymptotic behavior of $\lambda^\star_\mathrm{per}$ scales as
\begin{equation}\lambda^\star_\mathrm{per}(M,\delta) \sim 0.157(1.093)^M\lambda^\star = 0.157(1.093)^M\frac{\sqrt{\pi}(\pi M\delta)^{2M-1}(M-1)^{3/2}}{2^{4M-4}(2M-1)}.\end{equation}
In particular, much like the case of the real line superoscillations, the energy of the periodic superoscillations appear to be polynomial in the spacing of the prescribed points and exponential in the number of prescribed points. It must also be noted that the above asymptotic form was obtained by considering the behavior of superoscillations in which the bandlimit and the number of prescribed points were increased together. Therefore there is still some question about the behavior of the energy when the number of prescribed points are increased while keeping the bandlimit fixed (it should be noted that asymptotics for this case is not quite well-defined since the matrix $S$ will eventually become singular for a large number of prescribed points with a fixed bandlimit). However, by analogy with the real line case, we expect that the dependence of the energy on the bandlimit to be at most polynomial, which still suggests an exponential dependence on the number of prescribed points.

There appears to be strong numerical evidence for the above asymptotic form, but we do not yet have an analytic proof for the asymptotic behavior. Therefore the asymptotic form above should be regarded as a conjecture based on numerics, awaiting rigorous proof.  

\section{Conclusions}
We began by studying the properties of minimum-energy superoscillatory functions that are defined on the real line and are square integrable, focusing on their scaling behavior and overall shape. In particular, we found a curious interplay of extreme sensitivity with extreme stability: we showed that the superoscillatory stretch is extremely sensitive to perturbations to the generating coefficients, which translates into difficulties for realizations in the lab. But we also found that the overall shape of the superoscillatory function is very stable: the large-scale behavior of these functions tends to behave as a scalar multiple of a single superoscillatory function $\tilde{f}$ which only depends on the locations of the prescribed amplitudes. We identified the function $\tilde{f}$ as the minimum energy superoscillating function whose amplitudes at the prescription points are given by the coefficients of the eigenvector of the smallest eigenvalue of the prolate matrix.

Further, regarding the small scale behavior, i.e., regarding the details of the superoscillating stretch, we presented numerical evidence that the behavior of the superoscillatory stretch is very accurately approximated by a polynomial. In particular, for a superoscillatory function fitted through $N$ points, the superoscillatory stretch is very accurately approximated by the polynomial of degree $N-1$ fitted through the same points. This observation is summarized as a conjecture awaiting analytic proof. Under the conditions of the conjecture, an approximation for the Fourier transform of the superoscillatory stretch is presented.

We also generalized the construction of minimum-energy superoscillations to the case of periodic signals. We showed that in the periodic case, there exists a matrix $S$ which plays the role analogous to the prolate matrix $\rho$ for real line superoscillations. We presented numerical results for the energy growth of the periodic superoscillations and also conjectured an expression for their asymptotic behavior. An analytic proof for this asymptotic behavior is still needed however.\\

\noindent \bf Acknowledgments: \rm AK, LG and ET acknowledge support from the Discovery, Engage, and USRA programmes of the National Science and Engineering Research Council of Canada (NSERC), respectively.

\end{document}